\documentclass[11pt,a4paper]{article}
\usepackage{times,latexsym}
\usepackage{url}
\usepackage{amssymb}
\usepackage{amsmath}
\usepackage{multirow}

\usepackage[T1]{fontenc}

\usepackage[acceptedWithA]{tacl2021}
\usepackage{tacl2021}
\usepackage{booktabs}
\usepackage{amssymb}

\usepackage{tikz}
\usepackage{subcaption}
\usetikzlibrary{arrows.meta, positioning}

\definecolor{tokenblue}{RGB}{80,120,200}
\definecolor{tokenred}{RGB}{200,80,80}
\definecolor{tokengray}{RGB}{170,170,170}

\usepackage{xspace,mfirstuc,tabulary}

\newif\iftaclinstructions
\taclinstructionsfalse 
\iftaclinstructions
\renewcommand{\confidential}{}
\renewcommand{\anonsubtext}{(No author info supplied here, for consistency with
TACL-submission anonymization requirements)}
\newcommand{\instr}
\fi

\iftaclpubformat 

\else

\fi

\title{Beyond the Autoregressive Horizon: A Comprehensive Survey of Diffusion Models, World Modelling, and State Space Models for Code}

\author{
  Kishan Maharaj\thanks{\hspace{0.5em}These authors contributed equally.}
  \quad
  Ashita Saxena\footnotemark[1]
  \quad
  Srikanth Tamilselvam \\
  IBM \\
  \texttt{\{kishanmaharaj, ashitasaxena\}@ibm.com}\\
  \texttt{srikanth.tamilselvam@in.ibm.com}
}

\date{}

\begin{document}
\maketitle
\begin{abstract}
Autoregressive (AR) language models have driven significant progress in automated software engineering, enabling powerful code generation and assistance systems. However, the next-token prediction paradigm introduces structural limitations for code reasoning, including restricted global planning, challenges in maintaining long-range dependencies, and limited grounding in program execution semantics. Noting the heavy skewness of existing literature towards AR models, we discuss emerging paradigms that could potentially overcome the logic and scaling bottlenecks of next-token prediction by unlocking next-generation architectural capabilities for code intelligence.
Specifically, we discuss the potential of Diffusion Models, which generate code via holistic denoising that captures long-range syntactic constraints often missed by AR models. We also discuss Code World Models (CWMs), which simulate execution states to support reasoning, and State Space Models (SSMs), which provide linear-time efficiency for massive contexts. By connecting these developments with findings from cognitive neuroscience, we outline directions for developing "System 2" code generation agents.


\end{abstract}

\iftaclpubformat


\fi

\section{Introduction: The Autoregressive Bottleneck}

The dominant framework for modern Large Language Models (LLMs) is the autoregressive (AR) objective \cite{radford2018improving, radford2019language}, which models the joint probability of a sequence $x = (x_1,..., x_n)$ as the product of conditional probabilities: $P(x) = \prod_{t=1}^n P(x_t | x_{<t})$. While this formulation is mathematically tractable and highly effective for natural language, which flows linearly, aligning with human speech, it imposes a rigid sequentiality that can limit its expressiveness, which could lead to poor generalizability  \cite{chen2025coda, zhang2025exploring,lin2021limitations}.

The left-to-right generation process of AR models creates what we term the \textit{"Sequential Dependency Trap"}. In software engineering, early decisions (e.g., selecting a library version or defining a function/class) constrain code written later, while implementation details in function bodies may determine required imports at the beginning of the file, introducing dependencies that flow backwards relative to AR generation. When an AR model makes an early mistake, such as choosing a suboptimal data structure or misusing a function definition, it creates a "poisoned" context. As the model cannot backtrack, subsequent tokens must condition on this error, which cannot be corrected, leading to error propagation in which the model hallucinates justifications for the initial mistake \cite{zhang2023language, huang2025survey, kalai2025language}. 
\textit{In contrast, human developers code non-linearly: drafting function signatures, sketching implementations, adding missing imports, and iteratively refining the logic}. The "write-once" nature of AR inference, therefore, precludes this feedback loop.

Programming languages are formal languages governed by strict grammatical rules that often span long distances in code \cite{harrison1978introduction, allamanis2018survey}. For instance, a closing brace \} at line 500 must match an opening brace \{ at line 10, and a variable used deep within a loop must comply with type definitions established earlier in a header file. AR models rely on attention mechanisms \cite{vaswani2017attention} to reference prior tokens, yet they struggle to maintain such long-range dependencies precisely \cite{rando2025longcodebench, liu2025thus}, particularly as the context window becomes saturated with irrelevant information \cite{liu2024lost, shi2023distracted}. \textit{Although attention enables access to past tokens, it does not allow the model to revise earlier tokens to satisfy constraints that emerge later}. This limitation becomes pronounced in Fill-in-the-Middle (FIM) tasks \cite{bavarian2022efficient}, where generated code must remain consistent with both a prefix and a suffix. While AR models can be trained with FIM objectives \cite{wolf2023fill, fried2023incoder}, their architecture cannot natively treat the prefix and suffix as a jointly editable representation. Moreover, this training paradigm does not account for execution semantics that arise as interpreters or compilers process code sequentially.


Overall, traditional Autoregressive (AR) models face three key bottlenecks that limit their effectiveness in repository-scale software engineering. First, sequential dependency enforces unidirectional, token-by-token generation, preventing multi-token predictions across positions and disallowing modification of previously generated tokens. This irreversibility causes early mistakes to propagate, often resulting in cascading errors and hallucinations. Second, the computational cost of self-attention scales quadratically ($O(N^2)$), making the processing of large, repository-level contexts prohibitively expensive. Finally, AR models exhibit a semantic disconnect by treating code purely as text, lacking an inherent understanding of execution semantics. As a result, they may generate syntactically plausible but functionally incorrect logic that ignores the runtime behavior of the programming environment.

\begin{figure*}[!t]
    \centering
    \includegraphics[width=\textwidth]{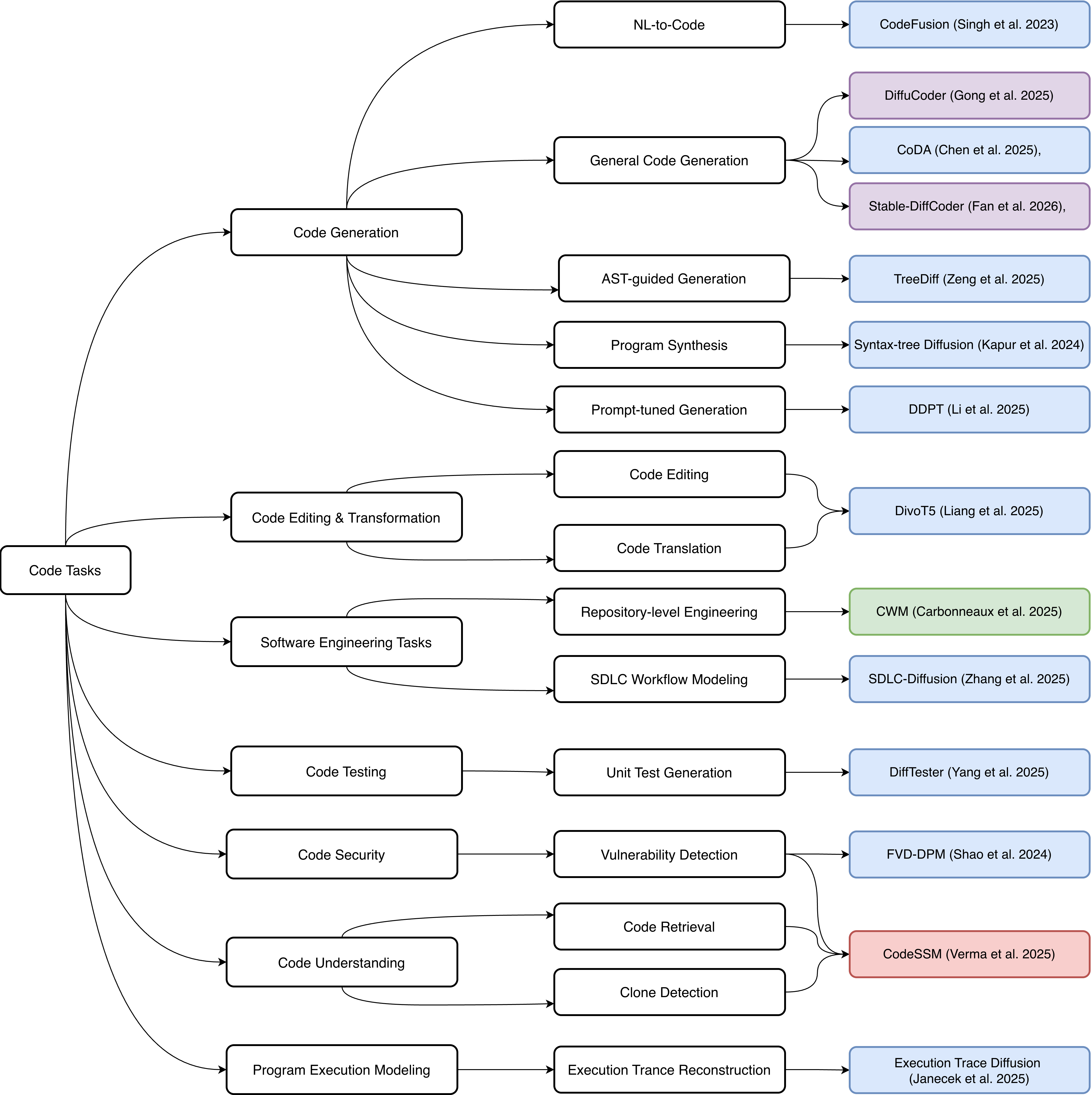}
    \caption{This diagram categorises non-autoregressive paradigms for different code-related tasks. Here, Blue represents diffusion-based approaches, green denotes World Modelling, and red corresponds to State Space paradigms. Based on this, we note that diffusion models remain the most explored non-autoregressive paradigm when compared to World Modelling and State Space Models.}
    \label{fig:taxonomy}
\end{figure*}

\begin{figure}[!t]
    \centering
    \includegraphics[width=0.45\textwidth]{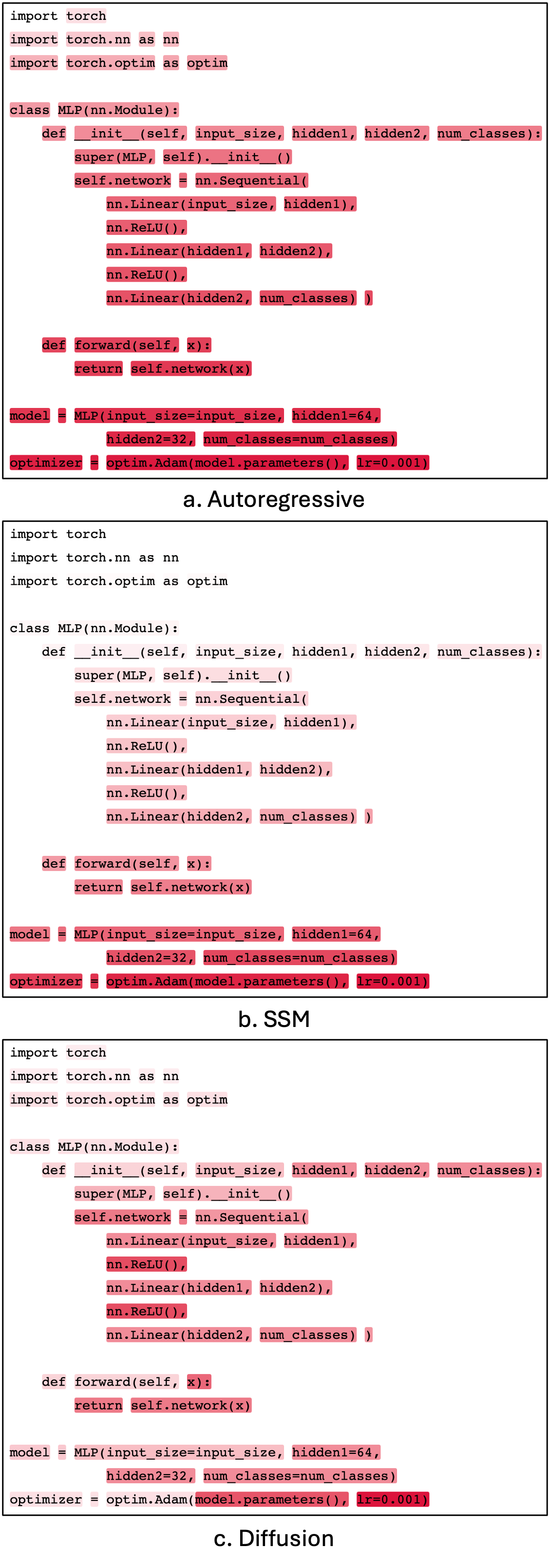}
    \caption{This diagram shows generation dynamics across paradigms. Here, intensity denotes token generation time. Unlike Diffusion’s asynchronous global refinement, Autoregressive models exhibit a steeper intensity gradient than SSMs, reflecting higher sequential decoding latency.}
    \label{fig:generation_comparison}
\end{figure}

While existing literature in automated software engineering remains largely centred on Autoregressive (AR) frameworks \cite{wang2025software, gu2025challenges, yang2025code, jelodar2025large, chen2025deep, li2025software, lyu2025automatic}, this work aims to broaden the discussion by examining emerging non-autoregressive paradigms relevant to code modelling (Figure \ref{fig:taxonomy}). Specifically, we analyse Diffusion Models, Code World Models (CWMs), and State Space Models (SSMs), highlighting how their architectural properties offer alternative mechanisms for code generation and reasoning (Figure \ref{fig:generation_comparison}). To the best of our knowledge, this work presents one of the first surveys that systematically discusses these paradigms within the context of code synthesis and software engineering. We hope this survey encourages further exploration of such approaches within the NLP and Software Engineering communities. We would like to further mention that this work is based on our understanding of open-source models, and we do not intend to comment or speculate on anything about closed-source models, which can potentially also have a component of non-autoregressive models (like diffusion, state space, etc.), for all we know. Our Contributions are:
\begin{itemize}

\item We present a structured survey of emerging non-autoregressive paradigms for code modeling, focusing on Diffusion Models, State Space Models (SSMs), and Code World Models (CWMs) by outlining their architectural principles and relevance to code tasks.

\item We conduct a comparative analysis of these paradigms, examining their trade-offs in terms of computational efficiency, scalability to long contexts, and structural consistency in generated code.

\item We connect these paradigms with insights from cognitive neuroscience, framing these paradigms within a perspective inspired by human “System 2” reasoning that emphasises deliberate analysis and error correction.

\end{itemize}
  
\section{Diffusion: The Paradigm of Iterative Refinement}

The emergence of Diffusion Models (DMs) for text generation \cite{sahoo2024simple, yi2024diffusion, nie2025large} and code synthesis \cite{xie2025dream, gong2025diffucoder, singh-etal-2023-codefusion} represents a fundamental architectural shift from autoregressive next-token prediction toward iterative refinement of the entire sequence.
Unlike AR models, which generate outputs sequentially and condition each step on previously generated tokens, diffusion-based approaches inspired by non-equilibrium thermodynamics \cite{sohl2015deep}, formulate generation as a denoising process that progressively transforms a noisy sequence into a coherent output. The growing maturity of this approach is reflected in recent large-scale and commercial diffusion-based systems, including Seed Diffusion \cite{song2025seed}, Mercury \cite{khanna2025mercury}, and Gemini Diffusion \cite{gemini_diffusion}, which demonstrate the viability of diffusion-based generation for long-form and high-fidelity text and code.  


\subsection{Foundation of Diffusion Models}

In this section, we discuss the foundations of diffusion models in the context of discrete data. Unlike image diffusion, which operates in continuous Gaussian space \cite{ho2020denoising, rombach2022high,chen2025comprehensive}, code generation requires handling discrete tokens \cite{shi2024simplified, sahoo2024simple, austin2021structured}. 

Recent advancements in discrete diffusion \cite{sahoo2024simple, shi2024simplified} have established a framework where the forward corruption process $q(\mathbf{x}_t | \mathbf{x}_{t-1})$ is characterised by an absorbing state, typically a special $[ \text{MASK} ]$ token. 

For a sequence $\mathbf{x}_0 = [w_1, w_2, \dots, w_L]$ of length $L$, the transition to a latent state at time $t \in [0, 1]$ is defined by an independent and identically distributed (i.i.d.) masking operation:

\begin{equation}
q(\mathbf{x}_t | \mathbf{x}_0) = \prod_{i=1}^L q(x_{t,i} | x_{0,i})    
\end{equation}

The transition probabilities for individual tokens are governed by a survival probability $\alpha_t$:

\begin{equation}
\begin{cases}
q(x_{t,i} = m \mid x_{0,i} = v) = 1 - \alpha_t, \\
q(x_{t,i} = v \mid x_{0,i} = v) = \alpha_t
\end{cases}
\end{equation}

where $\mathbf{x}_0$ represents the clean data, $\mathbf{x}_t$ is the corrupted state at time $t$, and $m$ denotes the absorbing $[\text{MASK}]$ state. The noise schedule $\alpha_t$ is a monotonically decreasing function such that $\alpha_0 = 1$ (clean data) and $\alpha_1 \approx 0$ (total corruption).

The generative process reverses this degradation by estimating the original tokens from the masked sequence by parameterising the above equation via a neural network (e.g., a Transformer) that learns the distribution $p_\theta(\mathbf{x}_0 | \mathbf{x}_t)$. To transition from a noisier state $t$ to a cleaner state $s$ ($s < t$), the reverse step is formulated as:

\begin{equation}
p_\theta(\mathbf{x}_s | \mathbf{x}_t) = \sum_{\hat{\mathbf{x}}_0} q(\mathbf{x}_s | \mathbf{x}_t, \hat{\mathbf{x}}_0) p_\theta(\hat{\mathbf{x}}_0 | \mathbf{x}_t)    
\end{equation}
Here, $\hat{\mathbf{x}}_0$ represents the model's categorical prediction over the vocabulary for all masked positions. In practice, the model $\mathbf{f}_\theta(\mathbf{x}_t, t)$ outputs logits which are converted to probabilities via a softmax layer.
The other simplified objective used for optimisation is a weighted cross-entropy loss focused exclusively on the masked tokens:

{\small
\begin{equation}
\begin{aligned}
\mathcal{L}(\theta)
&= \mathbb{E}_{\substack{
t \sim \mathcal{U}(0,1),\\
\mathbf{x}_0 \sim \mathcal{D},\\
\mathbf{x}_t \sim q(\mathbf{x}_t \mid \mathbf{x}_0)
}}
\Bigg[
\sum_{i \in \text{Mask}} \lambda(t)\,
\log p_\theta(x_{0,i} \mid \mathbf{x}_t)
\Bigg]
\end{aligned}
\end{equation}
}

where $t$ is sampled from a uniform distribution $\mathcal{U}(0,1)$, $\mathcal{M}$ denotes the set of indices where $x_{t,i} = [\text{MASK}]$, and $\lambda(t)$ is a time-dependent weighting factor derived from the noise schedule $\alpha_t$. This ensures the model prioritizes accurate reconstruction at various stages of the diffusion trajectory.

By training the network to predict the original uncorrupted tokens $x_0$ from a noisy state $x_t$, the model learns to capture global dependencies and structural constraints that sequential autoregressive models often overlook.


\subsection{Generation Dynamics in Diffusion}

Diffusion models fundamentally alter the generation dynamics by introducing \textbf{Parallel Decoding} and \textbf{Iterative Refinement}.

\textit{Parallel Decoding} enables the model to update all tokens in a sequence simultaneously rather than generating them one token at a time. This naturally supports multi-token prediction and bidirectional context awareness, allowing each token to be generated using both left and right context. For instance, a model may begin with a sequence of mask tokens (e.g., [MASK]) and predict likely locations of syntactic elements such as semicolons, function declarations, and return statements in a single step. This capability directly addresses limitations of purely left-to-right generation \cite{chen2025coda}.

\textit{Iterative Refinement} replaces single-pass decoding with $K$ denoising iterations, where the entire sequence is repeatedly updated. This allows the model to correct inconsistencies introduced in earlier steps, such as mismatches between variable definitions and later usage, which creates an inherent self-correcting mechanism. It also enables Step-Length Decoupling, where the number of denoising iterations is independent of sequence length; unlike autoregressive models, longer outputs do not require proportionally more decoding steps \cite{zhang2025exploring}.

Due to bidirectional context, these models hold the ability to satisfy fixed rule sets and formal grammars natively, something often overlooked in general NLP literature but critical for code \cite{cardei2025constrained}. Unlike autoregressive models, where enforcing constraints such as matching parentheses depends on the model probabilistically remembering earlier tokens \cite{mundler2025constrained}, diffusion models maintain the entire sequence at each timestep within a latent representation. This allows syntactic constraints to be optimised globally across the sequence \cite{singh-etal-2023-codefusion, xie2025dream}. For example, the presence of a closing brace at position $N$ can influence the generation of the corresponding opening brace at position $1$, since the model optimises the joint probability of the entire sequence rather than the conditional probability of the next token.

In practical scenarios, these features enable diffusion models to exhibit adaptive generative behaviours \cite{xie2025dream}, based on the complexity of the problem. 
For example, if the task requires a definite program structure, the model can establish the high-level program structure (e.g., imports, class definitions, main blocks) during early high-noise stages, effectively planning the code skeleton before filling in local details such as variable names or operations in later steps.
Conversely, for reasoning-intensive tasks (e.g., multi-step logic), generation may proceed in a non-linear, interleaved manner where key logical conditions are generated first, then supporting code, and finally refining edge cases, mirroring how solution structures often depend on salient insights.



This coarse-to-fine generation aligns with the hierarchical nature of code, where high-level architectural decisions constrain lower-level implementation details. It can be viewed as a generalisation of traditional left-to-right decoding used in autoregressive models, whose strictly sequential process often misaligns with real-world programming workflows \cite{li2025beyond, zhang2025exploring}. Consequently, these architectural freedoms enable global sequence-level reasoning and iterative correction during generation, reducing the error accumulation inherent to left-to-right decoding and offering a principled solution to the "Sequential Dependency Trap" of AR models.

\subsection{Diffusion Models for Software Development}

The above theoretical advantages have been harnessed in several recent works, which directly compare autoregressive models with diffusion models for code-related tasks \cite{zeng2025treediff, li2025ddpt, liang2025directional, janecek2025execution, yang2025difftester, kapur2024diffusion, singh2025diffusion}.

\citet{zhang2025exploring} explores the usage of diffusion models on the complete Software Development Lifecycle (SDLC), which includes code generation, defect detection, program repair, and cross-file maintenance. The study demonstrates, at a similar parameter scale (7B), DLLMs outperform AR-LLMs in terms of both accuracy and efficiency across six different benchmarking datasets. 


Beyond standard diffusion approaches, recent works \cite{xie2025dream, gong2025diffucoder, ye2025dream, chen2025coda} adapt autoregressive (AR) models into diffusion frameworks for code-related tasks \cite{gong2024scaling}. This adaptation relies on two key mechanisms. (1) Shift Operation: the target output is defined as the input sequence shifted left by one position, aligning the diffusion objective with the AR model’s pretrained tendency to predict the next token. (2) Attention Mask Annealing: the model gradually transitions from causal masking to full bidirectional attention. Instead of removing the causal constraint abruptly, the proportion of right-context tokens is progressively increased during training.
Addressing the high variance of token log-likelihood estimates and associated efficiency issues, \citet{gong2025diffucoder} further proposes coupled-GRPO, a sampling strategy that constructs complementary mask noise across training completions, thereby avoiding semi-autoregressive decoding.



\citet{chen2025coda} demonstrates the effectiveness of a progressive masking schedule that gradually increases training difficulty across epochs. Their model, CoDA (Coding via Diffusion Adaptation), is a 1.7B-parameter diffusion model trained on large-scale code corpora using this curriculum. Built on Qwen3-1.7B \cite{yang2025qwen3}, CoDA matches or surpasses diffusion models with up to 7B parameters while using roughly one quarter of the weights.
Block Diffusion \cite{arriola2025block} further combines autoregressive and diffusion paradigms by generating sequences block-by-block. Each block is produced via diffusion while conditioning on previously generated blocks, effectively performing autoregression at the block level. This design enables flexible-length generation and improved inference efficiency through KV caching and parallel token sampling. The approach is adopted in Stable-DiffCoder \cite{fan2026stable}, which builds on the Seed-Coder architecture and introduces a block-diffusion continual pretraining (CPT) stage with a tailored warmup and blockwise clipped noise schedule, resulting in improved benchmark performance.
By bridging AR and diffusion architectures, such hybrid models retain pretrained knowledge while gaining bidirectional context and parallel generation capabilities. Consequently, they achieve competitive results on major code benchmarks, including MBPP \cite{austin2021program}, HumanEval \cite{chen2021evaluating}, EvalPlus \cite{liu2023your}, and BigCodeBench \cite{zhuo2024bigcodebench}.

In the subsections below, we describe diffusion customised models specific to different tasks in software development. 

\subsubsection{Code Generation}

Various diffusion-based frameworks for code generation focus on integrating structural constraints into the diffusion process to ensure syntactic validity. 
The first diffusion-based NL-to-code model CODEFUSION \cite{singh-etal-2023-codefusion} combines an encoder-decoder architecture with a diffusion process by adapting continuous paragraph denoising (CPD) \cite{lin2023text} to code by only applying noise to tokens that correspond to identifiers in code or to built-in keywords in the target language. This enables the model to learn to restore these, focusing on code-critical tokens, leading to competitive results with auto-regressive code models and text diffusion models. 
Similarly, TreeDiff \cite{zeng2025treediff} operationalizes global constraint satisfaction by integrating the Abstract Syntax Tree (AST) directly into the diffusion process for code generation. By leveraging the hierarchical structure of the AST, it avoids the structural violations introduced by standard random masking and ensures that program reconstruction adheres to grammatical constraints while capturing long-range dependencies.
Along the similar lines, \citet{kapur2024diffusion} proposes neural diffusion models that operate on syntax trees of any context-free grammar for inverse graphics tasks. Here, the objective is a more niche code generation problem: converting images into programs that produce those images. Effectively, this framework extends the principles of image-based diffusion to the discrete, hierarchical structure of syntax trees, enabling the generation of programs that are both syntactically valid and visually representative of the target image.

Beyond direct generation, \citet{li2025ddpt} propose Diffusion-Driven Prompt Tuning (DDPT), which uses a diffusion model as an optimiser to generate high-performing prompt embeddings for code generation. Trained with a language modelling loss, the diffusion process synthesises task-specific embeddings from a standard Gaussian distribution. This approach addresses the "cold-start" issue of prompt initialisation, eliminating manual initialisation while maintaining a compact storage footprint for the learned diffusion weights.


\subsubsection{Code Editing and Test Case Generation}

Diffusion models are well-suited for code editing and repair due to their inherent denoising objective. \citet{singh2025diffusion} show that the final stages of diffusion closely resemble “last-mile repairs” \cite{bavishi2022neurosymbolic}, which target minor syntax errors or typo-level patches requiring minimal structural changes. By injecting controlled noise into a faulty snippet and resuming denoising, the model naturally performs localised and computationally efficient edits. This process also enables a powerful data augmentation strategy, where intermediate noisy states and final corrected programs sampled from the diffusion trajectory form large training datasets for specialised repair agents.

Complementing this, \citet{liang2025directional} introduces DivoT5, which models the step-by-step code editing behaviour of human developers using real-world editing trajectories. By framing the transition from a buggy or incomplete program to a correct version as a directed denoising process, the approach treats software evolution as a structured path toward a clean global state. This formulation yields strong performance across multiple code-editing benchmarks, including natural language–guided refinement, non-NL refinement, and bug fixing.




In the domain of quality assurance, \citet{yang2025difftester} proposes DiffTester, a diffusion-based approach for Unit Test Generation (UTG) that leverages the parallel generation capabilities of diffusion models. The authors identify a trade-off between efficiency and test quality, where generating more tokens per step can degrade test case quality. To address this, DiffTester exploits repetitive patterns in tests targeting the same focal method (the method whose functionality a unit test verifies). It parses generated code into multiple ASTs and extracts common nodes to guide generation. As a result, DiffTester achieves significant acceleration while preserving test coverage across three benchmarks and demonstrates strong cross-language generalizability.

\subsubsection{Security, Diagnostics, and System Analysis}

The diffusion paradigm also extends to fine-grained program analysis and system comprehension. \citet{shao2024fvd} propose FVD-DPM, a Conditional Diffusion Probabilistic Model that formulates vulnerability detection as a graph-structured prediction task. The approach constructs a fine-grained semantic representation of code and uses diffusion to model node label distributions in program slices, enabling the identification of vulnerable nodes.

Beyond vulnerability detection, \citet{janecek2025execution} explore diffusion-based time-series imputation models to reconstruct missing events in execution traces. These reconstructed traces support several software engineering applications, including program comprehension \cite{alanazi2021facilitating}, performance analysis \cite{noferesti2024enhancing}, fault diagnosis \cite{chen2021trace}, and anomaly detection \cite{janecek2022performance, panahandeh2024serviceanomaly}.

\section{State-Space Models for Code}
\label{sec:ssm-code}

\subsection{Foundations of State-Space Models}
\label{sec:ssm-foundations}

State-space models (SSMs) originate from classical control theory and
signal processing, where they are used to represent dynamical systems
whose internal state evolves over time in response to external inputs.
Recently, SSMs have been revisited as neural architectures for
efficient long-context sequence modeling, emerging as a promising
alternative to attention-based transformers
\citep{gu2020hippo, gu2021s4}.

An SSM processes a sequence by maintaining a latent state that
summarizes past information. Each input token updates this state,
which then produces the corresponding output representation. Unlike
transformers, which compute pairwise token interactions through
self-attention, SSMs propagate information forward through an evolving
hidden state.








Modern neural SSMs are commonly formulated as linear dynamical systems:

\[
\mathbf{h}_{t+1} = A\mathbf{h}_t + B\mathbf{x}_t
\]

\[
\mathbf{y}_t = C\mathbf{h}_t + D\mathbf{x}_t
\]

where $\mathbf{x}_t$ is the input at timestep $t$, $\mathbf{h}_t$ is a
latent state summarizing past context, and $\mathbf{y}_t$ is the output
representation. The matrices $A,B,C,$ and $D$ parameterize the state
transition and observation dynamics. In many neural SSM
implementations, the direct term $Dx_t$ is omitted or absorbed into
projection layers, resulting in the simplified form $y_t = Ch_t$.

A key theoretical foundation for modern SSMs is the HiPPO framework,
which compresses continuous signals into a fixed-dimensional state
representation using orthogonal polynomial projections
\citep{gu2020hippo}. Building on this idea, the S4 architecture
introduced structured state-space parameterizations that enable
efficient convolutional implementations and stable training on very
long sequences \citep{gu2021s4}. Subsequent work proposed simplified
variants such as diagonal state-space models (DSS/S4D)
\citep{gu2022dss} and S5 layers \citep{smith2022s5}, improving
scalability and integration into modern neural architectures.

A central advantage of SSMs is their computational complexity.
Transformers rely on self-attention with pairwise token interactions,
leading to quadratic scaling with sequence length. In contrast, SSMs
update the latent state once per token, resulting in linear complexity
and enabling efficient modeling of very long sequences.

\subsection{Efficient Long-Context Architectures}

The development of SSMs is part of a broader effort to address the
scalability limitations of transformer self-attention, whose quadratic
computational and memory costs become prohibitive for long documents,
program files, or repository-scale contexts.

Early work attempted to mitigate this limitation through sparse
attention mechanisms. Examples include Sparse Transformers
\citep{child2019sparse}, sliding-window attention in Longformer
\citep{beltagy2020longformer}, and block-sparse attention patterns in
BigBird \citep{zaheer2020bigbird}. Reformer further reduces attention
cost using locality-sensitive hashing to approximate token
interactions \citep{kitaev2020reformer}. Although these approaches
reduce computational cost, they still rely on attention as the
fundamental modeling operator.

Another line of work replaces attention with alternative sequence
modeling operators. Hyena models long-range dependencies using
implicit long convolutional filters \citep{poli2023hyena}, while
RetNet proposes a retention mechanism that approximates attention via
recurrent state updates \citep{sun2023retentive}. While these
architectures avoid quadratic attention costs, they often rely on
specialized parameterizations or training procedures.

State-space models provide a conceptually different solution by
modeling sequences as the evolution of a latent dynamical system. A
persistent hidden state summarizes the sequence history, enabling
information to propagate forward without explicitly attending to all
previous tokens. This yields linear computational complexity while
maintaining the ability to model long-range dependencies
\citep{gu2021s4, gu2023mamba}, making SSMs particularly well suited
for tasks involving extremely long contexts such as repository-scale
code understanding and execution trace modeling.

\subsection{Modern Neural State-Space Architectures}

Recent work has substantially improved the practicality of state-space models (SSMs) for large-scale sequence modeling. Early architectures such as S4 demonstrated strong performance on long-range sequence tasks but required specialized kernel implementations. Subsequent work introduced simplified parameterizations and improved initialization schemes that make SSM layers easier to integrate into modern neural networks \citep{gu2022efficiently, smith2023simplified}. A major breakthrough came with the selective state-space model \textsc{Mamba}, which enables input-dependent state transitions and hardware-efficient implementations \citep{gu2023mamba}. Unlike earlier SSMs with fixed transition dynamics, Mamba adapts its state updates based on the current input token, increasing expressivity while retaining linear-time inference; empirical results show that Mamba-based architectures can achieve language modeling performance comparable to transformers while scaling more efficiently to long contexts. Another emerging direction involves hybrid architectures that combine attention mechanisms with state-space layers, leveraging their complementary strengths: attention captures precise local token interactions, while SSM layers provide efficient long-range memory. Architectures such as Jamba integrate transformer attention blocks with Mamba layers to improve long-context reasoning in language models, and similar hybrid transformer--SSM designs are beginning to appear in large-scale industrial models for code and language understanding, suggesting that SSMs may ultimately serve as a complementary architectural component rather than a complete replacement for transformers.

\subsection{State-Space Models for Code}

Source code exhibits structural properties that differ from natural
language. Programs follow strict syntactic rules and contain
long-range dependencies that may span hundreds or thousands of
tokens. Variables defined early in a file may be referenced much
later, and function implementations often depend on imports declared
earlier in the program. Modeling such dependencies requires
architectures capable of preserving long-range contextual
information.

SSMs provide a natural architectural fit for these requirements. By
maintaining a persistent latent state summarizing previously observed
tokens, SSMs can capture global program context while scaling
linearly with sequence length \citep{gu2021s4, gu2023mamba}. This
makes them particularly attractive for repository-scale reasoning and
code analysis tasks.

Recent work has begun exploring SSMs for code intelligence problems.
The first explicit study in this direction is \textsc{CodeSSM}, which
evaluates SSM-based encoders on tasks such as code retrieval,
vulnerability detection, clone detection, and token classification
\citep{verma2025codessm}. Experimental results indicate that
SSM-based models can achieve competitive or superior sample
efficiency compared to transformer baselines while scaling to longer
contexts with lower memory overhead.

Recent analysis has also examined how SSM-based models represent
program semantics internally. For example, \citet{wu2026ssmcode}
Investigate the representations learned by SSM models during
pretraining and fine-tuning, showing that while these models capture
syntactic and semantic properties of code effectively, maintaining
long-range dependencies during downstream adaptation remains an open
challenge.

Overall, the application of SSMs to code remains an emerging research
area. Most existing work focuses on code understanding tasks rather
than full program synthesis or automated software engineering
workflows. Nevertheless, the ability of SSMs to efficiently model
long contexts suggests that they may play an important role in future
code language models, particularly in hybrid transformer--SSM
architectures.

\paragraph{State-Space Models in Non-Autoregressive Generation.}

While most applications of state-space models in code have focused on
autoregressive or encoder-style architectures, recent work explores their
integration with non-autoregressive generative frameworks such as diffusion
models. Structured State Space Diffusion (SSSD) models incorporate SSM
backbones (e.g., S4) into score-based diffusion processes for sequential
generation and imputation \citep{alcaraz2023sssd, tashiro2021csdi}. Variants
such as SSSDS4 and SSSDSA replace convolutional or attention-based diffusion
backbones with state-space sequence models, enabling efficient modeling of
long-range dependencies during iterative denoising.

By combining holistic sequence reconstruction with linear-time memory,
these models offer an alternative to next-token prediction that enforces
global consistency across long contexts. Although primarily studied for
time-series data, they have direct implications for code intelligence.
Execution traces and program state transitions can be viewed as structured
temporal sequences, where missing segments correspond to partial observations
of program behavior. Recent work applies diffusion-based models to execution
trace reconstruction, showing that SSM-based variants can effectively recover
long-range behavioral patterns \citep{hamou2025trace}.
\paragraph{From State-Space Models to World Models for Code.}

An important conceptual connection between state-space models and
recent world-model approaches for code lies in their shared emphasis
on \emph{stateful dynamics}. While SSMs model sequences through the
evolution of a latent state summarizing past inputs, world models
learn transitions between program states induced by code execution or
agent actions. Execution traces such as variable updates,
control-flow transitions, and environment interactions can therefore
be viewed as partial observations of an underlying program state.
This perspective suggests a natural convergence: SSMs provide an
efficient backbone for modeling long execution traces, while
world-model pretraining grounds these states in executable semantics.

\section{Code World Modelling for Learning Execution Semantics}

While contemporary LLMs primarily model next-token prediction, World Models aim to simulate next-world predictions, capturing possible future states of an environment under consideration \cite{xing2025critiques}. This capability relates to a model’s ability to simulate multiple potential outcomes and reason about which trajectory is most likely to achieve a desired goal. In psychological terminology, this process is often referred to as hypothetical thinking \cite{ball2020hypothetical}, which is analogous to the ability to mentally simulate future scenarios. Such foresight can enable predictive models to exhibit stronger reasoning capabilities by allowing them to anticipate environmental dynamics and make decisions in novel situations.

For example, astronauts are trained for scuba diving to develop an understanding of how their bodies move under altered physical conditions, often useful when adapting to environments such as low gravity during lunar exploration \cite{weiss2012simulation}. Analogously, if models are exposed to states of a coding environment across diverse inputs or execution traces, they may learn to anticipate the consequences of different reasoning paths by simulating alternative program executions.


A fundamental critique of current LLMs is their treatment of code as text through next-token prediction, which models statistical correlations between tokens rather than the causal dynamics of the code. While this objective can approximate aspects of world modelling in text-based environments \cite{li2025word}, it does not explicitly encode information about code execution. Predicting code linearly, left-to-right and top-to-bottom, enables strong syntactic pattern recognition but may fail to capture program execution semantics. Consequently, models often generate code that appears correct (syntactically valid) yet fails during execution due to logical errors or runtime crashes \cite{carbonneaux2025cwm}. Mastering coding, therefore, requires understanding not only what code looks like but also what it does when executed, motivating training mechanisms that incorporate execution-aware signals.



Building on these ideas, Code World Models (CWMs) \cite{carbonneaux2025cwm} aim to simulate the code execution environment within a neural architecture. The key intuition is that by modelling how the execution of a line of code changes the state of local variables (within the stack frame), a system can reason about how these changes propagate through the program and ultimately affect its outputs. In this framework, the objective is to learn execution semantics by simulating how updates to local variables ripple across the codebase, thereby improving awareness of real runtime behaviour and enabling more robust code generation. Formally, a world model takes the current world state $s$ and an action $a$ as input, and predicts the next world state $s'$ via a transformation function, typically modeled as a conditional probability distribution:

\[
s' \sim p(s' \mid s, a)
\]



Within a code environment, the training objective involves two key components: \textit{execution traces} and \textit{agentic interaction traces}. Execution tracing captures the state of a Python program at intermediate execution points corresponding to events in the Python interpreter by running executable units (e.g., functions or repository images) with different input–output pairs or CI tests while recording the state of memory. Agentic interaction tracing, in contrast, models the interaction between an LLM-based software engineering agent and a computational environment within the SWE-bench framework \cite{jimenez2023swe}, collecting multi-step trajectories generated when an LLM is prompted to solve a software engineering task in the context of a specific code repository.



Predicting execution traces enables a model to learn the execution semantics of programs by simulating observation–action dynamics within a computational environment, thereby aligning generated code with the corresponding program state at each step of execution. Complementarily, learning from agentic interaction traces exposes the model to the complexities of real-world software engineering by training on multi-step trajectories within the SWE-bench framework. Through these interactions, the model learns how actions such as editing files or running tests affect the broader repository state, providing an early grounding in environment dynamics that improves the efficiency of subsequent post-training.


The importance of these training mechanisms is empirically validated through ablation studies showing that code world modelling data, such as Python execution traces and executable Docker environments, can directly improve downstream task performance. These findings highlight that for an LLM to excel at software engineering, it must do more than predict the next token; it must model the causal relationship between code and its execution.

\section{Cross-Paradigm Benchmarking Trends}
Table~\ref{tab:benchmark-empirical} summarizes representative empirical
evaluations across major modeling paradigms for code. Rather than
presenting a unified evaluation landscape, the table highlights how
different paradigms are assessed using distinct tasks, benchmarks, and
metrics.

Autoregressive models remain closely tied to classical code generation benchmarks such as HumanEval and APPS. These evaluations focus on functional correctness of short programs using pass@k metrics, which align naturally with next-token prediction and one-shot program synthesis. As a result, much of the empirical progress reported for autoregressive systems is concentrated on snippet-level code generation.

Diffusion-based approaches exhibit a more diverse evaluation profile. Across the works listed in Table~\ref{tab:benchmark-empirical}, diffusion models are evaluated not only on natural language to code generation tasks such as CoNaLa, but also on structured generation, code repair, editing, testing, and software engineering workflows. Correspondingly, their evaluation metrics extend beyond pass@k to include execution match, structural similarity, coverage metrics, and workflow-level performance indicators. This diversity reflects the iterative generation process of diffusion models, which enables global sequence refinement and structured modifications.

State-space models appear in the literature primarily as representation learners rather than generative models. Existing evaluations focus on code understanding tasks including retrieval, vulnerability detection, and clone detection, using metrics such as MRR, F1, and accuracy. This suggests that the generative capabilities of SSM-based architectures for program synthesis remain largely unexplored.

World-model approaches represent a further shift in evaluation scope. Instead of isolated code snippets, these systems are evaluated on repository-scale software engineering benchmarks such as SWE-bench, Terminal-Bench, and CruxEval. These tasks measure success using execution-grounded metrics such as resolve rate and task completion, capturing the ability of models to interact with development environments and perform multi-step reasoning.

Taken together, the table reveals that benchmarking practices have evolved alongside modeling paradigms. Autoregressive models dominate snippet-level code generation benchmarks, diffusion models are increasingly evaluated on structured and iterative development tasks, state-space models are assessed primarily through representation-based benchmarks, and world models shift evaluation toward realistic software engineering workflows. While this diversification reflects the expanding scope of code intelligence research, it also complicates direct empirical comparison across paradigms and highlights the need for more unified evaluation frameworks capable of measuring long-context reasoning, iterative editing, and execution-aware software engineering.

\begin{table*}[!t]
\centering
\footnotesize
\setlength{\tabcolsep}{4pt}
\renewcommand{\arraystretch}{1.2}

\begin{tabular}{p{2.2cm} p{3.2cm} p{3.0cm} p{3.2cm} p{2.2cm}}
\toprule
\textbf{Paradigm} &
\textbf{Model / Paper} &
\textbf{Task} &
\textbf{Benchmark} &
\textbf{Metric} \\
\midrule

Autoregressive &
Codex \citep{chen2021evaluating} &
Code generation &
HumanEval, APPS &
pass@k \\

\midrule

Diffusion &
CodeFusion \citep{singh-etal-2023-codefusion} &
NL-to-Code generation &
CoNaLa, Bash, CF &
Template match, Execution match \\

Diffusion &
TreeDiff \citep{zeng2025treediff} &
Code generation (AST-guided) &
HumanEval(+), MBPP(+) &
pass@1 \\

Hybrid Diffusion &
DiffuCoder \citep{gong2025diffucoder} &
Code generation &
HumanEval, MBPP, BigCodeBench, EvalPlus &
pass@k \\

Diffusion &
SDLC-Diffusion \citep{zhang2025exploring} &
Software engineering tasks &
SDLC benchmark suite &
pass@k, workflow metrics \\

Diffusion &
DivoT5 \citep{liang2025directional} &
Code editing, translation &
CodeReview, CodeTrans &
EM, BLEU, CodeBLEU \\

Diffusion &
CoDA \citep{chen2025coda} &
Code generation &
HumanEval, MBPP, EvalPlus &
pass@1 \\

Diffusion &
Stable-DiffCoder \citep{fan2026stable} &
General coding tasks &
HumanEval(+), MBPP(+), BigCodeBench, LiveCodeBench &
pass@1 \\

Diffusion &
DDPT \citep{li2025ddpt} &
Prompt tuning for generation &
CodeAlpaca, CoNaLa &
BLEU, CodeBLEU \\

Diffusion &
DiffTester \citep{yang2025difftester} &
Unit test generation &
TestEval benchmark &
Coverage, throughput \\

Diffusion &
FVD-DPM \citep{shao2024fvd} &
Vulnerability detection &
NVD, SARD datasets &
F1, AUC, MCC \\

Diffusion &
Execution Trace Diffusion \citep{janecek2025execution} &
Execution trace reconstruction &
Phoronix test suite &
Accuracy, ROUGE-L \\

Diffusion &
Syntax-tree Diffusion \citep{kapur2024diffusion} &
Program synthesis &
TinySVG, CSG2D &
IoU \\

\midrule

State-Space Models &
CodeSSM \citep{verma2025codessm} &
Code understanding (retrieval, vulnerability detection, clone detection) &
NLCodeSearch, Devign, DiverseVul &
MRR, F1, Accuracy \\

\midrule

World Models &
CWM \citep{carbonneaux2025cwm} &
Repository-level software engineering &
SWE-bench Verified, Terminal-Bench, CruxEval &
Resolve rate, pass@1 \\

\bottomrule
\end{tabular}
\caption{Benchmarking practices across major modeling paradigms for
code. The table summarizes representative works, tasks, datasets, and
metrics used to evaluate autoregressive, diffusion-based, state-space,
and world-model approaches.}
\label{tab:benchmark-empirical}

\end{table*}

\section{Discussion: Cognitive Alignment of Code Models}

In this section, we discuss insights from cognitive neuroscience to inform the architecture of intelligent coding agents while highlighting the missing components in AR models. To build truly intelligent coding systems, it is essential to look at the only known example of general coding intelligence: the human brain.


A common misconception in autoregressive (AR) pre-training is treating programming strictly as a natural language task, i.e., predicting the next token. While this approach successfully mimics linear human speech, it fundamentally diverges from how humans write code. Coding is rarely a forward-only, linear process; it intrinsically involves iterative revision, debugging, and structural refinement. By restricting generation to a forward-only pass without the freedom to revisit and edit past tokens, pure AR models remain cognitively misaligned with the actual task of programming \cite{li2025beyond}. Furthermore, reducing code to a simple language modelling task forfeits the encoding mechanisms necessary to ground the model in the execution environment and understand the causal effects of the code.


This misalignment is supported by recent fMRI studies on programming students \cite{ivanova2020comprehension}, which reveal that code comprehension does not primarily recruit the brain's language centres (such as Broca's and Wernicke's areas). Instead, it heavily activates the Multiple-Demand (MD) Network, specifically the independently localised, left-lateralised frontoparietal regions associated with logical reasoning, mathematical problem-solving, and executive control. This aligns with the findings of \citet{mahowald2024dissociating}, who argue that the neural regions (within the brain) responsible for reasoning are functionally distinct from those governing language processing.

Building on this, evidence points to System 2 Recycling: the brain "recycles" evolutionary mechanisms originally adapted for complex tool usage and logic to handle programming, rather than relying on canonical linguistic circuits. For instance, when an expert programmer processes a recursive function, the brain maps the recursive stack onto the same neurobiological substrates used for planning multi-step mechanical tasks and hierarchical tool-use sequences \cite{ivanova2020comprehension}. Consequently, the logic of recursion functions as a mental instrument for structural state-tracking rather than a traditional grammatical construction.


These findings necessitate a paradigm shift in AI architectures. Specifically, they motivate the development of Code World Models that simulate the execution environment in a latent space, treating functions as causal "tools" that manipulate program states. Furthermore, the top-down planning characteristic of pre-motor logic suggests that Diffusion Models offer a promising path forward. Unlike AR models, diffusion generates code through global structural refinement, mirroring human cognitive planning.

The cognitive process of debugging and refining code bears a striking resemblance to the diffusion process. Code editing is inherently evolutionary; it is rarely a one-off task but requires gradually modifying relevant snippets to meet shifting requirements \cite{liang2025directional}. This incremental modification aligns perfectly with step-by-step denoising. When a programmer drafts code, they typically begin with a vague mental model (high noise) and iteratively "denoise" it by fixing syntax, clarifying variables, and resolving logic bugs \cite{gong2025diffucoder}. This "Draft-Review-Refine" loop is the core of human programming, and diffusion models formalise it. The reverse diffusion process ($x_T \to x_0$) can be conceptualised as a "trajectory of thought," where a high-entropy intent is progressively sharpened into a low-entropy, valid program. Simultaneously, as a programmer traces local variable changes and their rippling effects across the codebase, they are essentially utilizing an internal Code World Model.

A complementary capability required for programming is the ability to
maintain and update a persistent representation of program state while
processing long code sequences. When programmers reason about code, they
continuously track variable values, control flow, and intermediate states
across many lines of execution. This behaviour resembles the working-memory
processes supported by the Multiple-Demand (MD) network, which is responsible
for maintaining and manipulating task-relevant state during complex reasoning
\citep{duncan2010multiple, baddeley2012working}. Architectures based on
state-space models (SSMs) offer a computational abstraction of this mechanism
by maintaining a compact latent state that evolves over time as new tokens are
processed \citep{gu2021s4, gu2023mamba}. This evolving state is consistent with
the dynamical systems perspective of neural computation, where cognitive
processes can be interpreted as trajectories over latent states
\citep{wang2008decision, mante2013context}, enabling efficient tracking of
long-range dependencies and execution dynamics. Consequently, SSMs provide a
promising architectural backbone for modeling extended reasoning trajectories
in code.

Therefore, to properly align an AI framework with the neural substrates of human programming, we must explicitly model the environment state, logical refinement, and hierarchical structure of code. These insights suggest that the next generation of code intelligence will be defined by hybrid architectures that synthesise the strengths of multiple paradigms to simulate this human approach grounded in cognitive principles. 

In the existing scenario, it would be ideal to leverage the robust post-training innovations of AR models while integrating the unique capabilities of alternative architectures. For example, Block Diffusion
\cite{arriola2025block} successfully combines AR and diffusion models by sequentially generating blocks of tokens, applying diffusion-based generation within each block while conditioning on previously generated ones.

Ultimately, a cognitively aligned architecture could synthesise several complementary paradigms. Diffusion Models can enable global planning and ``edit-in-place'' refactoring through iterative refinement, while Code World Models grounded in AR foundations ensure that code is treated as
an executable tool with causal effects on program state. State-space models (SSMs) can further provide an efficient mechanism for maintaining persistent representations of program state across long contexts, supporting reasoning over large files and execution traces. Finally, incorporating a System 2 reasoning loop using post-training mechanisms
such as Reinforcement Learning with Verifiable Rewards (RLVR)
\cite{shao2024deepseekmath} can allow these agents to plan, verify, and execute software tasks with human-level reliability.

\section{Conclusion and Future Scope}

In this survey, we examined emerging non-autoregressive paradigms for code intelligence and their implications for automated software engineering. While autoregressive next-token prediction has driven substantial progress in code large language models, it imposes structural limitations on complex tasks. Unlike natural language, programming is inherently non-linear; developers routinely revise prior code, reason about program behaviour, and plan across extensive contexts. Consequently, treating code as purely sequential text neglects its stateful, executable nature. Emerging paradigms, such as diffusion models, state-space models (SSMs), and code world models, mark a crucial shift from token-level prediction toward program-level reasoning, offering a promising foundation for next-generation code models.

We further observe that while diffusion-based methods have reached a degree of maturity in code tasks, SSMs and world modelling remain in an exploratory phase. Looking forward, the most promising direction may lie in hybrid architectures that combine the strengths of multiple paradigms. Rather than replacing autoregressive models and their existing innovation, future systems can benifit by building upon their generative capabilities while incorporating mechanisms for global refinement, long-context reasoning, and execution grounding. Early examples such as Block Diffusion \cite{arriola2025block} illustrate this direction by combining autoregressive generation with diffusion-based refinement. Together, these advances suggest a broader transition toward planning, iterative editing, and execution-aware reasoning in next-generation code intelligence systems.

\section{Limitations}

A key limitation of this survey is the difficulty of making direct empirical comparisons across modeling paradigms. Existing works evaluate autoregressive models, diffusion models, state-space models, and world-model-based systems on different tasks, datasets, and evaluation metrics. This fragmentation of benchmarking practices makes it challenging to perform consistent cross-paradigm comparisons on a single standardized benchmark

\bibliography{tacl2021}
\bibliographystyle{acl_natbib}

\appendix

\iftaclpubformat

\onecolumn

\end{document}